\documentclass[runningheads]{llncs}

\usepackage[T1]{fontenc}
\usepackage{graphicx}
\usepackage{amsmath,amssymb}
\usepackage{url}
\usepackage{hyperref}
\usepackage{svg}
\usepackage{subcaption}
\usepackage{booktabs}
\usepackage{orcidlink}
\usepackage[misc]{ifsym}

\title{Shadows of Transparency: \\ Signaling Transparency-Impeding Behavior Using Public Data \thanks{Accepted manuscript on May 5, 2026, to the 25th IFIP International Conference on Electronic Government (EGOV 2026).}}
\titlerunning{Shadows of Transparency}

\author{Jos Zuijderwijk\textsuperscript{(\Letter)}\orcidlink{0009-0008-9561-9354} \and
Iris Beerepoot\orcidlink{0000-0002-6301-9329} \and
Thomas Martens\orcidlink{0009-0007-2353-7301} \and
Eva Knies\orcidlink{0000-0003-0918-7572} \and
Tanja van der Lippe\orcidlink{0000-0002-5245-4659} \and
Hajo A. Reijers\orcidlink{0000-0001-9634-5852}}

\authorrunning{Zuijderwijk et al.}

\institute{Utrecht University, Princetonplein 5, 3584 CC Utrecht, The Netherlands\\
\email{a.j.h.zuijderwijk@uu.nl} \and
Ministry of Infrastructure and Water Management, The Hague, The Netherlands}

\begin{document}
\maketitle
\begin{abstract}
Freedom of Information (FOI) laws aim to increase government transparency, yet existing assessments focus mainly on legal compliance and procedural outcomes, leaving organizational behavior underexamined. As FOI processes are increasingly mediated through digital information systems, public response data offer traces of how organizations handle requests and disclosures in practice. This paper develops and evaluates a pattern-based approach that uses such data to signal transparency-impeding behavior across government agencies and over time. Drawing on prior literature, we identify nine recurring behavioral patterns that undermine transparency. Using a dataset of 15,277 Dutch FOI dossiers comprising 139,449 documents, we operationalize and evaluate measurable indicators for six of these patterns, demonstrating that they enable systematic comparison of disclosure practices across agencies and over time. Expert interviews with researchers, journalists, and civil servants confirm the interpretability and practical usefulness of the indicators for investigative prioritization, comparative research, and transparency oversight.
\end{abstract}

\section{Introduction}
\label{sec:introduction}

Government transparency is widely recognized as one of the pillars of democratic governance. It enables citizens to observe and evaluate their government by providing access to information about its decision-making process and, consequently, to hold their government accountable \cite{Hood2006}. This principle has been institutionalized through Freedom of Information (FOI) legislation, which grants citizens the right to request government information and obliges public agencies to respond within defined timeframes. FOI laws, often described as the backbone of transparency \cite{Grimmelikhuijsen2012}, have been adopted in almost all democratic countries \cite{FreedominfoND}.

Yet the mere existence of FOI laws alone does not ensure transparency in practice (e.g., \cite{Trautendorfer2025}). The performance of FOI legislation is typically assessed through field studies analyzing variables that influence request outcomes, such as approval rates or processing delays (e.g., \cite{Worthy2016,Rodriguez2018,Jenkins2020,Spac2025}). FOI performance is usually assessed at a procedural level, for example, using response times and exemption rates (e.g., \cite{Hazell2010,Wasike2020}), or at a legal level, considering the scope of exemptions or the presence of enforcement mechanisms (e.g., \cite{Ackerman2006,Mueller2019}). Such metrics capture formal FOI compliance but reveal little about how organizations behave when handling requests. This matters, because transparency is shaped in the day-to-day handling of records: e.g., how requests are interpreted or how searches are conducted, which exemptions are repeatedly invoked, how heavily documents are redacted, in what format information is released, and when disclosure is made. Consider, for example, the Dutch childcare benefits scandal, in which parents received government files so heavily redacted that they could not understand the basis for decisions affecting them \cite{NOS2019}. The agency was formally compliant---it responded to requests and applied legal exemptions---yet the disclosed information was rendered meaningless. Prior work suggests that such transparency-impeding behavior occurs within government agencies \cite{Pasquier2007,Luscombe2017}, but is difficult to observe directly and systematically across organizations and over time. Existing FOI assessment approaches mainly measure compliance, but they offer limited means to make such recurring transparency-impeding behavior visible.

The rise of e-government has created new empirical opportunities to study this otherwise hidden behavior through digital data. Information systems are now central to all aspects of managing public information \cite{GilGarcia2018,Kautto2020}, and FOI procedures are no exception \cite{Stratton2023}. Governments worldwide have launched digital platforms for managing FOI requests and releasing the requested information\footnote{Examples include \url{https://whatdotheyknow.com} (UK), \url{https://fragdenstaat.de} (Germany), \url{https://righttoknow.org.au} (Australia), and \url{https://asktheeu.org} (EU). All accessed September 2025.}. As FOI requests are increasingly processed, logged, and published through information systems, they generate data that indirectly reflect disclosure practices. For instance, an agency that consistently releases documents as scanned images rather than searchable digital files makes information difficult to find and analyze. This would leave observable traces in public data as low machine-readability scores. Although FOI response data remain underused in empirical research, a few recent works have begun to explore them with regard to responsiveness, legislative differences, and the topics of citizen requests \cite{Trautendorfer2024,Ruohonen2025,Trautendorfer2025,Zuffova2025}.

Building on this observation, this paper introduces a pattern-based approach for signaling transparency-impeding behavior in the FOI process using public response data. By identifying and operationalizing recurring behavioral patterns that impede transparency, we show how disclosure practices can be systematically compared through publicly available FOI data. A similar pattern-based approach has recently been used to study government transparency through logs of document management systems \cite{Zuijderwijk2025}. These internal logs, however, are usually not accessible to researchers. In contrast, our study applies the approach to publicly available FOI response data, thereby extending its applicability to large-scale transparency research.

Accordingly, our research question is: \textit{How can recurring patterns in public FOI response data be operationalized into indicators of transparency-impeding behavior that allow for comparisons across government agencies and over time?}

We answer this question using a large dataset of Dutch FOI responses. We identify recurring behavioral patterns from existing literature, implement measurable indicators to examine technical feasibility and comparative outputs, and evaluate their interpretability and practical usefulness through expert interviews with researchers, journalists, and civil servants. In doing so, the paper contributes a pattern-based approach that complements compliance-oriented FOI assessment by enabling comparison of transparency-impeding behavior through publicly available data.

\section{Research Background}
\label{sec:background}

This section establishes the elements necessary for our approach. We first examine how organizational behavior can impede transparency. We then discuss how the FOI process is mediated through information systems.

\subsection{Transparency and its Impediments}
\label{sec:transparency}

Transparency is a broad term. In general, it can be described as ``looking through the window of an institution to see its inner workings'' \cite{denBoer1998}. In terms of information relations, transparency can be defined as the availability of information about an actor that enables other actors to monitor the performance or inner workings of that actor \cite{Grimmelikhuijsen2012,Meijer2013}. Following this definition, one could infer that maximizing transparency means disclosing as much information as possible that is relevant to public decision-making. There are, however, legitimate reasons to constrain the release of government information, such as national security concerns, privacy protection, or commercial confidentiality \cite{Pasquier2007}.

Apart from legitimate reasons for withholding information, governments may be reluctant to disclose relevant information despite legal requirements for transparency \cite{Henninger2017}. It has been argued that governments can maintain FOI laws while loosening the constraints imposed by those laws through less visible administrative action, thereby creating the appearance of openness while limiting disclosure in practice \cite{Roberts2006}.

Building on the concept of transparency as the availability of information that enables one actor to observe another's inner workings, we can distinguish between internal and external transparency \cite{Street2004}. \textit{Internal} transparency refers to an organization's ability to know what information exists, where it is stored, and how it can be retrieved and interpreted. It reflects the quality of internal information management, i.e., how records are created, organized, and maintained within information systems. \textit{External} transparency, by contrast, denotes the degree to which internally available information is made visible to actors outside the organization, e.g., citizens or journalists.

For the present study, a further distinction is useful. Formal FOI compliance concerns whether agencies meet legal and procedural requirements, for example regarding timeliness or the formal use of exemptions. Disclosure behavior concerns recurring organizational practices in how requests are interpreted, how records are selected, how information is redacted, how documents are formatted, and when disclosures are made. The argument of this paper is that existing FOI assessment primarily captures formal compliance, while disclosure behavior remains comparatively difficult to observe even though it shapes transparency in practice.

In the age of e-government, external transparency increasingly takes the form of digital information disclosure. Whereas open data platforms have been widely studied as tools for proactive disclosure of data (e.g., \cite{Matheus2023}), FOI platforms supporting reactive disclosure have received comparatively little scholarly attention. The next section examines the FOI process and its implementation within information systems.

\subsection{FOI Process Within Information Systems}
\label{sec:foi_process}

\begin{figure}[t]
  \centering
  \includegraphics[width=0.6\textwidth]{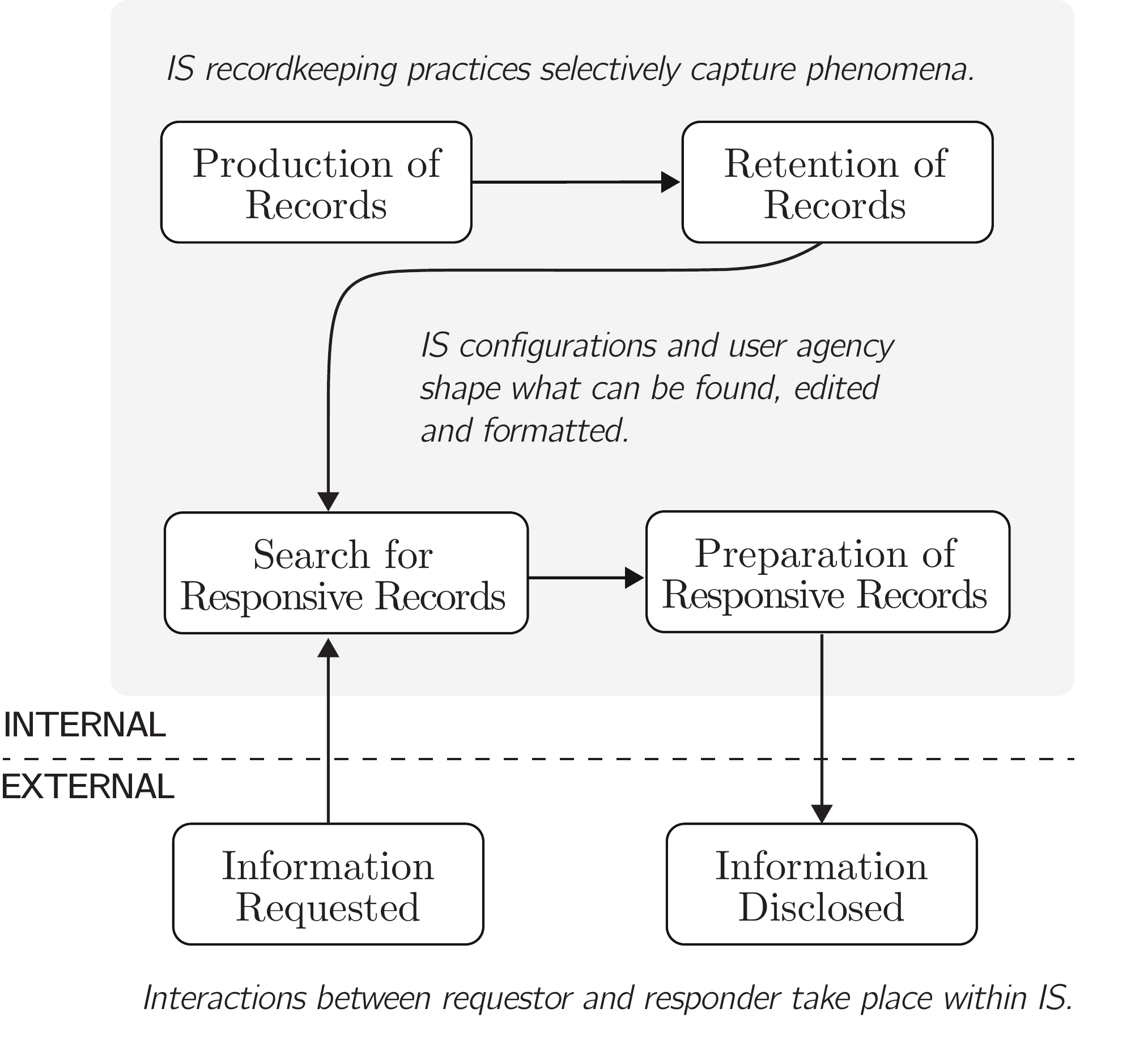}
  \caption{FOI process through information systems (IS) from request to disclosure. Adapted from Carter and Stratton \cite{Carter2021}. The internal/external boundary (dashed line) and minor wording adjustments have been added by the authors.}
  \label{fig:foi}
\end{figure}

FOI regimes generally combine two modes of disclosure: reactive and proactive. \textit{Reactive} (or passive) disclosure refers to the FOI process in which individuals submit requests for information, agencies locate the relevant records, apply any necessary exemptions, and then release the material. In contrast, \textit{proactive} (or active) disclosure requires public agencies to routinely publish certain categories of information on their own initiative, without waiting for a specific request \cite{Ruijer2017}. This study focuses on reactive disclosure, as we are interested in request-response behavior. Throughout the paper, the term `FOI process' refers to this reactive mode.

Information systems play a mediating role throughout the FOI process \cite{Carter2021}. As shown in Figure~\ref{fig:foi}, they influence each step of the reactive FOI process, from the submission of a request to the disclosure of information. The process begins when information is requested by an external actor, triggering internal information system activities. Within the organization, records are first produced and subsequently retained according to recordkeeping practices, which determine what aspects of phenomena are captured in the first place. When an FOI request arrives, civil servants use information systems to search for responsive records, i.e., records within the scope of the request, relying on the system's structure and search capabilities to locate relevant material. The identified records are then prepared for release, a stage where system configurations and user agency influence how documents are redacted, edited, and formatted. Finally, information is disclosed externally, completing the process.

The internal and external dimensions are interdependent. Internal transparency is a prerequisite for external transparency: information that cannot be located or understood internally cannot be disclosed externally, regardless of legal requirements (cf.\ strained transparency \cite{Pasquier2007}). Conversely, external transparency influences internal information management. Knowing that records might become public later, civil servants may document more, or less, carefully.

\section{Research Approach}
\label{sec:approach}
While data on internal information management, such as system logs or recordkeeping metadata, is rarely available to researchers, data from the external side of the FOI process is public by definition. Our study aims to complement existing transparency assessments using public data. As argued in Section~\ref{sec:background}, transparency is shaped by organizational disclosure behavior, i.e., the concrete actions and decisions of civil servants processing FOI requests. This behavior is mediated by information systems and can leave recurring traces in public response data.

To answer the research question, we identify recurring behavioral patterns from the literature, map them onto observable features in public FOI datasets, operationalize one or more indicators for each pattern, and apply them to a Dutch dataset to examine whether they enable comparison of disclosure practices across ministries and over time. We then evaluate the interpretability of these indicators and the practical usefulness of our method through expert interviews.

\subsection{Behavioral Patterns}
\label{sec:approach_patterns}

Transparency is a concept that is multi-faceted and inherently difficult to measure directly. We therefore construct proxies at the behavioral level of civil servants within organizations. Because it is more straightforward to identify behavioral patterns that signal impeded transparency than those that reflect its successful achievement, we focus on recurring organizational practices within the reactive FOI process that undermine transparency. We refer to behavioral patterns of transparency-impeding behavior as repeated but counterproductive routines. A similar approach to identifying recurring problematic behavior has been applied in the study of internal government information management \cite{Zuijderwijk2025}. Our study extends this logic to externally visible FOI disclosure data. For the present study, we scope such patterns to behavior within the FOI process as mediated through information systems and reflected in public response data. We derive behavioral patterns from prior literature and subsequently evaluate them with domain experts.

\subsection{Public Data and Indicators}
\label{sec:approach_indicators}

From the behavioral patterns, we move to the empirical level, where public data from FOI platforms are analyzed. An appropriate dataset for this purpose contains at least partial FOI dossiers, associated metadata such as request and response dates or file types, and sufficient organizational or temporal coverage to support comparison. FOI response platforms and datasets derived from them typically provide case-level data on the handling of requests, document-level metadata on the disclosed files, and content-level information from the documents themselves. Together, these elements capture both procedural and substantive aspects of disclosure. Where content-level data are not directly available, they can often be extracted from linked files through OCR. Building on the behavioral patterns and these data, we derive a set of indicators. These indicators do not prove the existence of transparency-impeding behavior; rather, they signal patterns that may warrant further scrutiny. Higher indicator values suggest a greater prevalence of the corresponding behavioral pattern. Once defined, the indicators can be computed per dossier and aggregated at appropriate analytical levels such as ministry or year, enabling longitudinal and cross-organizational comparison.

\subsection{Evaluation}
\label{sec:approach_evaluation}

Because the indicators are proxies derived from metadata rather than direct observations of behavior, evaluation is needed to establish whether they are interpretable as signals of the behavioral patterns they are intended to capture. In this study, evaluation does not establish whether a ministry acted strategically in a particular case. Rather, it examines whether the indicators are technically feasible, interpretable, and practically useful as signals of transparency-impeding behavior. A key aspect of this evaluation is whether the indicators enable comparison across agencies and over time, as specified in our research question.

\section{Case Study}
\label{sec:case}

We apply our approach to Dutch FOI responses to demonstrate how transparency-impeding behavior can be signaled and compared across ministries. The goal of the case is not to make definitive claims about ministries' actual intent or misconduct, but to show how indicators derived from public FOI data can enable systematic comparison of disclosure practices within a single government.

\subsection{Behavioral Patterns}
\label{sec:case_patterns}

For the purpose of our case study, we identified a set of behavioral patterns by drawing on existing literature. We scoped the analysis to recurring behavior within the reactive FOI process that may impede transparency and that can leave observable traces in public response data. We exclude internal information-management problems not observable through public responses and non-behavioral impediments such as high FOI fees \cite{Wagner2023} or broader infrastructural constraints \cite{Pasquier2007}. The patterns were derived through a scoped review of literature on FOI administration and obstructive disclosure practices, using \cite{Luscombe2017} as a starting point and extending through snowballing. Nine behavioral patterns were identified, summarized in Table~\ref{tab:antipatterns}.

\begin{table}[t]
\centering
\caption{Identified behavioral patterns of transparency-impeding behavior in the FOI process.}
\scriptsize
\begin{tabular}{@{}p{3.6cm}p{7.2cm}l@{}}
\toprule
Behavioral Pattern & Description & Source\\
\midrule
Obstructive Formatting & Releasing information in forms that hinder search and analysis, e.g., scanned printouts. & \cite{Luscombe2017,Tong2025,Meijer2002}\\[3pt]
Excessive Redaction & Obscuring text so extensively that disclosed records lose meaning and context. & \cite{Luscombe2017}\\[3pt]
Catch-All Exemption & Repeated reliance on broad exemption grounds beyond their intended scope. & \cite{Luscombe2017,Zuffova2023,Arnold2014}\\[3pt]
Narrow Reading & Interpreting requests restrictively to reduce what must be disclosed. & \cite{Luscombe2017}\\[3pt]
Overbroad Reading & Interpreting requests so expansively that key information is buried in volume. & \cite{Luscombe2017}\\[3pt]
Stale Release & Delaying disclosure long enough to reduce its value or impact. & \cite{Treadwell2016}\\[3pt]
Renamed Units & Relabeling terms or units after exposure so that subsequent requests yield no results. & \cite{Luscombe2017}\\[3pt]
Document Suppression & Reducing or ceasing production of certain records after prior exposure through FOI. & \cite{Luscombe2017}\\[3pt]
Reclassification & Shifting records into more restrictive categories to make them harder to access. & \cite{Pasquier2007,Galison2004}\\
\bottomrule
\end{tabular}
\label{tab:antipatterns}
\end{table}
Obstructive Formatting can be illustrated by releasing scanned printouts rather than machine-readable files, which limits not just readability but also analytical use of disclosed information (cf.\ \cite{Meijer2002} on informational vs.\ analytical transparency). Excessive Redaction concerns heavily redacted documents. For Catch-All Exemptions, provisions such as ministerial secrecy may be stretched beyond their intended scope \cite{Zuffova2023,Arnold2014}. Narrow Reading and Overbroad Reading concern how requests are interpreted: one reduces the scope of disclosure, the other buries relevant information in volume. Stale Release concerns timing: delaying disclosure diminishes the value of the released information. Renamed Units, Reclassification, and Document Suppression all concern disappearance from the disclosure record, whether through relabeling, reclassifying sensitivity, or ceasing documentation.

\subsection{Public Data}
\label{sec:case_data}

We selected the Woogle\footnote{Woogle is now called WooZM, see \url{https://woozm.nl}} dataset \cite{vanHeusden2025} to operationalize the behavioral patterns identified above where feasible. The Woogle dataset was constructed through web scraping of Dutch government FOI platforms. The dataset contains responses from all Dutch ministries, provinces, and 31 of 342 municipalities, spanning the years 2001--2025. The dataset includes both reactively and proactively released files. For this study, we filtered the dataset on the reactively released documents to focus on behavior within the request-response process. The filtered dataset contains 15,277 dossiers, 139,449 documents, and 2,644,696 pages.

\begin{figure}[t]
    \centering
    \begin{subfigure}[b]{0.42\textwidth}
        \fbox{\includegraphics[width=\textwidth]{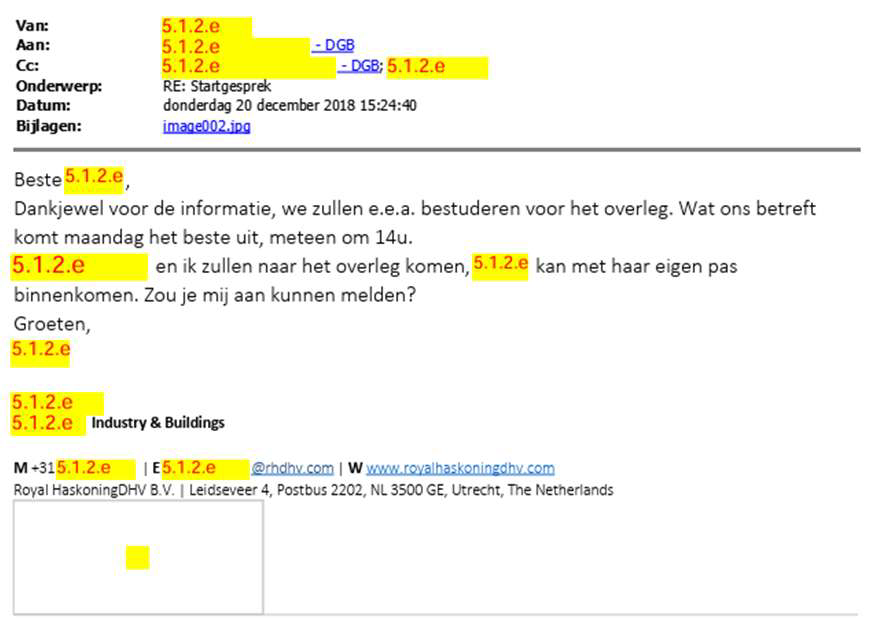}}
    \end{subfigure}
    \hfill
    \begin{subfigure}[b]{0.55\textwidth}
        \fbox{\includegraphics[width=\textwidth]{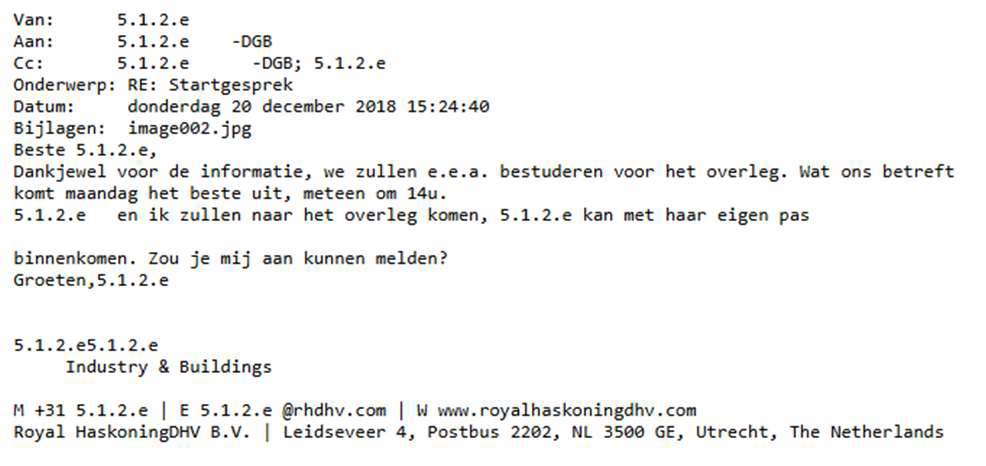}}
    \end{subfigure}
    \caption{Excerpt from an example FOI dossier: a released email and its corresponding OCR text. For reference: \url{https://pid.wooverheid.nl/?pid=nl.mnre1130.2i.2023.14}}
    \label{fig:example_foi}
\end{figure}

The dataset combines three interconnected data components: case-level data on request handling (request details, decision texts, organizational identifiers), document-level metadata on disclosed files (file characteristics, creation timestamps, format specifications), and page-level data (OCR results, redaction measurements, accessibility scores). An FOI dossier consists of a request, a response, an inventory list, and the disclosed documents themselves. Figure~\ref{fig:example_foi} shows an example of a released document from the dataset. Redacted portions are highlighted with yellow markers (though marker styles vary greatly across documents), with the legal basis\footnote{Under the Dutch FOI Act (Wet open overheid), exemption grounds are specific legal provisions (e.g., Article 5.1.2.e for personal privacy) that justify withholding information. Some grounds are absolute (always block disclosure) while others are relative (require a balancing test).} for each redaction noted inside the marked areas.

\subsection{Indicators and Results}
\label{sec:case_results}

\begin{figure}[t]
  \centering
  \includegraphics[width=\textwidth]{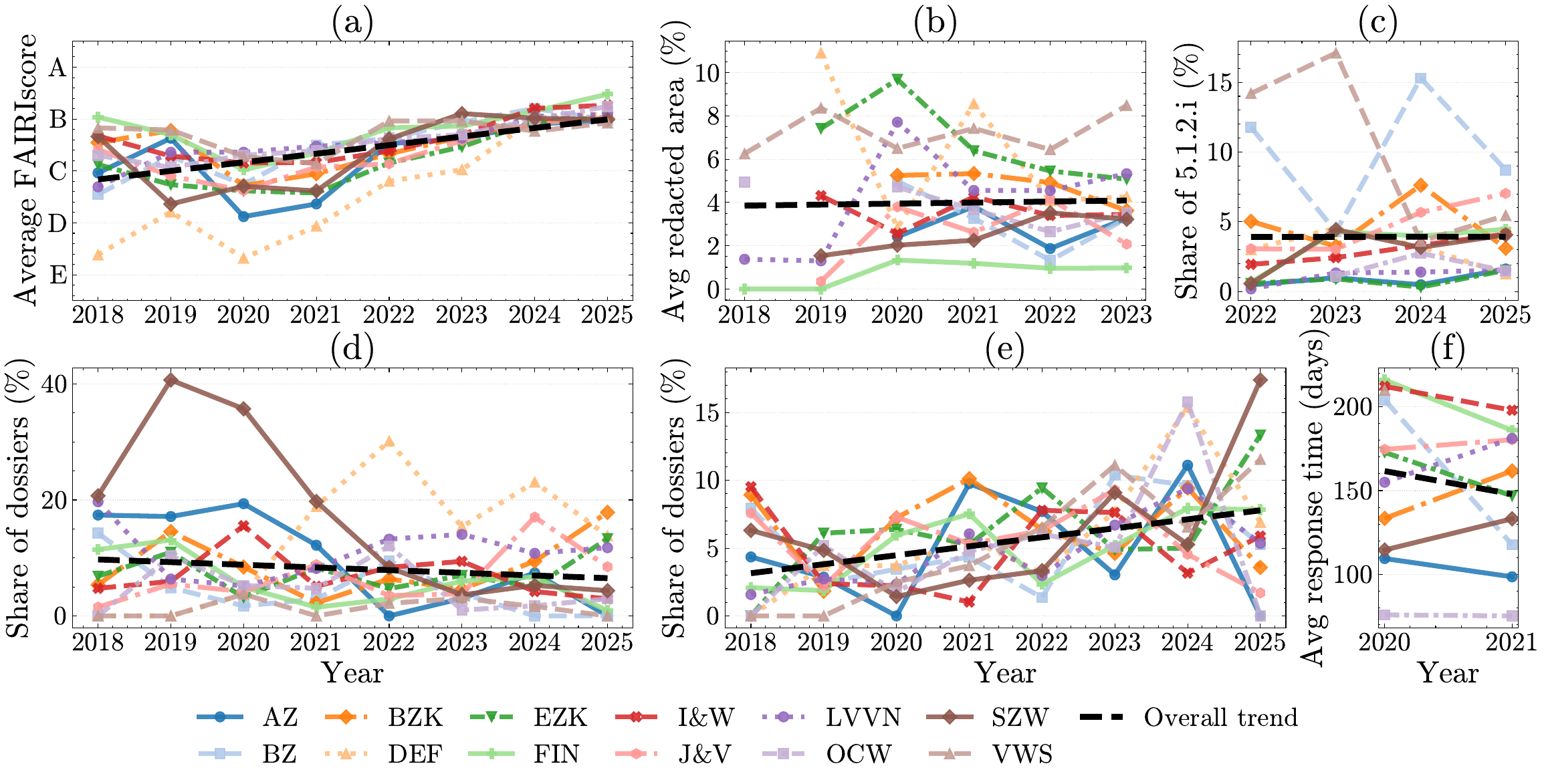}
  \caption{Six indicators of transparency-impeding behavior per ministry over time. Top row: (a) Average FAIRIscore (Obstructive Formatting; A = best, E = worst); (b) Redaction rate (Excessive Redaction); (c) Catch-all exemption share (Catch-All Exemption; share of exemption ground 5.1.2.i). Bottom row: (d) Low-volume dossier rate (Narrow Reading); (e) High-volume dossier rate (Overbroad Reading); (f) Average response time in days (Stale Release). Dashed black line indicates the overall trend.
}
  \label{fig:results}
\end{figure}

We expressed each of the nine behavioral patterns identified in Section~\ref{sec:case_patterns} as measurable indicators based on observable characteristics in the Woogle dataset. As described in Section~\ref{sec:approach_indicators}, indicators are continuous measurements that signal the presence of a behavioral pattern. Of the nine patterns, we could operationalize six using the metadata, redaction measurements, and accessibility scores available in the dataset: Obstructive Formatting, Excessive Redaction, Catch-All Exemption, Narrow Reading, Overbroad Reading, and Stale Release. The remaining three---Renamed Units, Document Suppression, and Reclassification---require semantic topic modeling over time, which exceeded the scope of the present study; they are discussed further in Section~\ref{sec:discussion}. We implemented the six indicators in Python. Data availability varies by indicator: Stale Release could be computed only for 2020–2021 due to missing timestamps, Catch-All Exemption from 2022 onward following the new FOI Act, Excessive Redaction for 2018–2023 based on available redaction measurements, and the remaining indicators for 2018–2025. Three ministries established in 2024 were omitted from the figures. We refer to ministries by their official abbreviations.\footnote{AZ (General Affairs), BZ (Foreign Affairs), BZK (Interior), DEF (Defence), EZK (Economic Affairs), FIN (Finance), I\&W (Infrastructure), J\&V (Justice), LVVN (Agriculture), OCW (Education), SZW (Social Affairs), VWS (Health). Full names and English translations are included in the repository (see Data and Code Availability).}

\textbf{Obstructive Formatting} is operationalized as the \textit{average FAIRIscore}, defined as the mean FAIRIscore \cite{Marx2023} across all documents within an organization per year. The FAIRIscore grades machine readability and accessibility on a scale from A (best) to E (worst); scores closer to E capture formats that hinder text extraction, such as scans. Figure~\ref{fig:results}a shows the results. The overall trend improves clearly, moving from predominantly C- to D-level scores in 2018 toward B-level by 2024--2025, with the dashed overall trend line capturing this shift. DEF and AZ show the most pronounced early dips, with DEF recording the lowest
formatting scores around 2019--2020 before both recover to B-level. Cross-ministerial differences persist throughout the period, but the convergence toward B-level suggests a broad shift in formatting practices, consistent with the 2022 Dutch FOI law's mandate for accessible document formats \cite{Drahmann2021}.

\textbf{Excessive Redaction} is operationalized as the \textit{redaction rate}, defined as the average percentage of text area redacted, detected algorithmically \cite{vanHeusden2023}. Figure~\ref{fig:results}b shows the results, covering the period 2018--2023. EZK shows the most pronounced peaks, reaching approximately 9\% in 2020, well above the overall average of approximately 4\%. VWS also shows elevated rates. FIN and J\&V remain consistently below the overall average throughout the period. The overall trend is relatively stable, suggesting neither a tightening nor loosening of redaction practices across ministries. The substantial cross-ministerial variation, with DEF showing notable year-to-year volatility, raises questions about whether redaction practices are proportionate to the sensitivity of the material disclosed.

\textbf{Catch-All Exemption} is operationalized as the \textit{catch-all exemption share}, defined as the proportion of exemption ground 5.1.2.i (``proper functioning of the State'') among all cited exemption grounds per ministry and year. Prior literature identifies this provision as one that can function as a catch-all in the Dutch FOI context due to its broad and discretionary applicability \cite{Drahmann2024}. Exemption grounds were detected by matching each document's OCR text against a
curated list of code variants (e.g., \texttt{5.1.2.i}, ``5.1 lid 2 sub i'') using regular expressions. Figure~\ref{fig:results}c shows the results, available from 2022 onward. While the overall average remains around 4--5\%, VWS shows a pronounced peak exceeding 15\% in 2023, and BZK and BZ show elevated use before declining. 

\textbf{Narrow Reading and Overbroad Reading} are operationalized as two indicators using dossier size as a proxy for how broadly a request has been interpreted. The \textit{low-volume dossier rate} is defined as the proportion of dossiers whose page count falls at or below the 5th percentile of the pooled dossier-size distribution across all ministries (two pages or fewer), signaling potential restrictive interpretation. The \textit{high-volume dossier rate} is the proportion at or above the 95th percentile (636 pages or more), signaling potential overbroad interpretation. Global thresholds ensure comparability across ministries. Figure~\ref{fig:results}d--e show the results. The low-volume dossier rate shows SZW spiking around 2019-2020 before declining sharply. AZ shows consistently elevated rates in 2018--2019, remaining above 15\%, and DEF spikes to roughly 30\% in 2021. The high-volume dossier rate is comparatively rare across all ministries and shows no clear outliers, with values generally remaining below 15\%. Both indicators differentiate ministries less consistently over time than the indicators discussed above.

\textbf{Stale Release} is operationalized as the \textit{average response time}, defined as the mean number of days between the submission of a request and the publication of the corresponding dossier. Due to missing timestamps, this indicator could only be computed for 2020 and 2021. DEF is absent from panel~(f) only, as it has no recorded 2020--2021 request dates. Figure~\ref{fig:results}f shows the results. Despite this limited window, the data reveal substantial cross-ministerial variation: average response times range from approximately 80 to over 200 days. FIN and I\&W show the longest average response times, approaching or exceeding 200 days, while OCW shows among the shortest. The overall trend slightly decreases between the two years, though several ministries show an upward trend. When disclosure takes months, the information may lose its relevance for public debate, journalistic investigation, or policy scrutiny.

Taken together, the six indicators show that public FOI response data can surface cross-ministerial and temporal variation in transparency-impeding behavior. Catch-All Exemption, Excessive Redaction, Obstructive Formatting, and Stale Release generate the clearest comparative signals. The Narrow Reading and Overbroad Reading indicators reveal some outliers, e.g., SZW's spike in low-volume dossiers, but differentiate ministries less consistently over time. Notably, certain ministries (e.g., VWS) appear as outliers across multiple indicators simultaneously, suggesting that transparency-impeding patterns may cluster within particular organizations.

\subsection{Evaluation}
\label{sec:evaluation}

To evaluate our approach, we conducted six semi-structured expert interviews, each lasting approximately one hour. Participants were selected to capture different forms of expertise: two academic researchers (one full professor specializing in transparency, one PhD researcher involved with the dataset), two investigative journalists specializing in FOI, and two civil servants with an FOI-related function from a Dutch ministry. All participants were shown the set of behavioral patterns, the corresponding indicators, and the outputs from the Dutch case. The interview protocol combined
a shared evaluative core with role-specific variants tailored to participants' expertise. Across versions, the protocol addressed three overarching dimensions: feasibility, quality and usability. Interviews were audio-recorded with consent and transcribed. The full protocol is provided in the repository. Responses were grouped thematically across participants along the three dimensions. The findings below were obtained by grouping recurring observations across the six transcripts along the three dimensions; we report points raised by multiple participants or flagged as particularly salient.

Implementation of the approach on Dutch FOI response data showed that several behavioral patterns can be operationalized and applied to a large dataset. Indicators such as dossier volume rates, redaction rates, and FAIRIscores were successfully implemented, demonstrating technical feasibility. Participants noted that the remaining patterns (Renamed Units, Document Suppression, and Reclassification) are inherently difficult to measure, because published FOI data provide visibility only into what is disclosed. While data quality is improving as more documents are digitally born and systematically published, one participant highlighted recurring sources of error such as OCR inconsistencies, divergent file structures, and incomplete metadata.

Participants widely recognized the conceptual plausibility of the identified behavioral patterns, judging them to reflect real practices in Dutch FOI processing. Catch-All Exemption and Excessive Redaction were identified as very common. Narrow Reading was described as a common but negotiated practice, whereas Renamed Units and Overbroad Reading were seen as rare or non-existent in the Dutch context.

A central question in our evaluation was whether the indicators enable comparison across agencies and over time. When presented with the indicator outputs, participants were able to derive specific comparative interpretations. For instance, the high redaction rates at certain ministries were linked by participants to the sensitivity of policy domains those ministries cover, such as health or economic affairs, while the improving FAIRIscores were attributed to the 2022 legislative mandate for accessible document formats. The exemption concentration indicator prompted participants to distinguish between expected dominance of the privacy exemption (5.1.2.e), which they considered unremarkable, and the frequent use of the ``proper functioning of the State'' exemption (5.1.2.i), which several experts flagged as a potential marker of catch-all behavior (related to the frequency of other exemption grounds). Critically, participants emphasized that the indicators' value lies precisely in comparison: a single ministry's score is hard to interpret in isolation, but relative differences and temporal trends raise questions about why one ministry behaves differently from another.

Additionally, participants identified behavioral patterns not included in the current set, including the offshoring of records to private messaging channels (e.g., WhatsApp, Signal) and procedural discouragement tactics (e.g., characterizing requests as abusive or excessively burdensome). Related behaviors, such as avoiding documentation, simply declaring that records do not exist, and masking departments, were likewise reported as recurrent in practice. Each suggests potential measurable proxies, although participants stressed that such indicators remain highly context-dependent.

All participants agreed that the approach holds practical value for enabling systematic comparison. Journalists saw it as a means to support investigative prioritization by identifying which agencies warrant closer scrutiny. Civil servants viewed it as a diagnostic tool for internal comparison, noting that ``dashboards could show why one ministry still does what another stopped doing''. Researchers emphasized its potential for comparative, longitudinal analysis of transparency and its contribution to the empirical study of digital government. Participants agreed on two main guidelines for interpretation. First, indicators should serve as diagnostic signals, not definitive proof of intent: ``they should raise questions, not point fingers''. Second, indicators need to be considered alongside context to avoid misreading them. Several participants proposed combining quantitative indicators with qualitative follow-ups. For example, one participant noted a structural shift in recordkeeping: ``there are many more documents now, and the types of documents have changed,'' with chat and messaging apps supplementing traditional memos and making disclosure ``much more practically difficult''.

\section{Discussion}
\label{sec:discussion}

This study set out to examine how recurring patterns in public FOI response data can be operationalized into indicators of transparency-impeding behavior that allow for comparisons across government agencies and over time. The results suggest that this is feasible for several behavioral patterns. In particular, Catch-All Exemption, Excessive Redaction, Obstructive Formatting, and Stale Release produced indicators that enable cross-ministerial and temporal comparison. Expert interviews confirmed that these indicators are interpretable and practically useful as diagnostic signals: they reflect recognized practices, surface variation that experts could contextualize, and were seen as valuable for journalism, research, and internal oversight alike.

A consistent finding across indicators is that cross-ministerial variation is substantial, while temporal trends differ by indicator. The gradual improvement in FAIRIscores, for instance, suggests that legislative mandates can have observable effects on disclosure behavior, whereas redaction levels remain broadly stable, indicating neither a tightening nor loosening of disclosure boundaries. Not all indicators performed equally well, however. The Narrow Reading and Overbroad Reading indicators differentiated ministries less sharply, likely because dossier size is shaped by many factors beyond request interpretation, including the nature of the topic and the volume of relevant records. This suggests that proxy-based indicators work best when the observable feature is closely tied to the behavioral pattern, as with redaction rates or exemption shares, and less well when the mapping is indirect.

The paper contributes in two ways. First, it contributes conceptually by distinguishing between formal compliance, disclosure behavior, and transparency, and by arguing that disclosure behavior forms an analytically important but insufficiently visible layer between the first and the third. Second, it contributes methodologically by showing how public FOI response data can be used as a source for pattern-based analysis of recurring transparency-impeding behavior across organizations and over time. In this sense, the paper complements existing FOI assessments. The indicators should be interpreted as signals that help prioritize further investigation.

The proposed approach is expected to be applicable beyond the Dutch context. Several jurisdictions, including Germany, the United Kingdom, and Australia, maintain public FOI platforms capable of supporting comparable analyses. Cross-national comparisons could reveal whether behavioral patterns follow similar trajectories across administrative cultures or reflect country-specific disclosure norms. Moreover, linking behavioral indicators to established metrics of FOI performance \cite{Hazell2010} or broader transparency assessments \cite{Williams2015} may help validate the results further.

Our study also has limitations. Because we rely on publicly released FOI responses, our analyses are constrained by the quality and completeness of the available data. Some data, such as request dates, were incomplete. Three behavioral patterns---Renamed Units, Document Suppression, and Reclassification---remain unoperationalized. These patterns require content-level analysis to detect whether topics vanish and reappear under new names. Their implementation is left for future work. For the Obstructive Formatting indicator, the FAIRIscore is an ordinal measure (A--E), which constrains quantitative analysis, and it is subject to recalibration over time, meaning longitudinal comparisons may partly reflect changes in the scoring instrument; reporting the share of D- and E-rated documents is one alternative operationalization that may mitigate the latter issue. More broadly, the current implementation focuses on demonstrating that systematic comparison is achievable; alternative operationalizations, e.g., using medians instead of means to reduce sensitivity to outliers, may better reflect disclosure behavior in practice. Future work can extend the implementation with more advanced techniques such as topic modeling \cite{Eickhoff2017} and refined exemption detection, expand the catalogue of behavioral patterns, and examine the content of FOI requests in more detail, although original request texts are not always publicly available.

\section{Conclusion}
\label{sec:conclusion}

This study introduced a pattern-based approach for signaling transparency-impeding behavior in the FOI process using publicly available response data. Drawing on existing literature, we identified nine recurring behavioral patterns that undermine transparency and operationalized six of them as measurable indicators. Applying the approach to Dutch FOI dossiers revealed variation across ministries and over time: formatting practices improved following legislative change, redaction levels varied substantially across policy domains, reliance on broad exemption grounds proved unevenly distributed, and response times differed markedly between agencies.

Expert interviews with journalists, researchers, and civil servants confirmed the interpretability and practical usefulness of the indicators, and their comparative value, while highlighting the need for careful contextual interpretation. Intent is not captured in the data, and our approach does not replace legal or qualitative evaluation. Rather, it provides a scalable, data-driven mechanism to signal behavioral patterns that may warrant closer scrutiny.

Ultimately, the findings suggest that public FOI response data can serve to signal recurring disclosure behavior that may shape transparency. As governments continue to digitize the FOI process, this approach offers a foundation for comparative research, transparency monitoring, and informed oversight.

\subsubsection*{Data and Code Availability.}
Our code and the interview protocol are available at \url{https://edu.nl/dwbr3}.

\bibliographystyle{splncs04}
\bibliography{references}

\end{document}